\documentclass[3p]{elsarticle}

\usepackage{hyperref}

\usepackage{amssymb}
\usepackage{bm}
\usepackage{dcolumn}
\usepackage{amssymb}
\usepackage{amsmath}
\usepackage{graphicx}

\usepackage{epstopdf}
\usepackage{epsfig}
\usepackage{xcolor}
\usepackage{color}
\usepackage{subfig}
\usepackage{booktabs}
\usepackage{multirow}
\usepackage{setspace}
\usepackage{array}
\usepackage{threeparttable}
\usepackage{txfonts}

\usepackage{exscale}
\usepackage{relsize}

\usepackage{ulem}
\usepackage{xcolor}

\newcommand{\new}[1]{{\color{black}{#1}}}

\journal{arXiv, published as: H.~Song \& B.-Q.~Ma, \href{https://doi.org/10.1016/j.dark.2025.101808}{Phys.Dark Univ. 47 (2025) 101808.}}

\begin{document}
\renewcommand{\arraystretch}{1.3}

\begin{frontmatter}

\title{Examining Lorentz invariance violation with three remarkable GRB photons}

\newcommand*{\PKU}{School of Physics, Peking University, Beijing 100871, China}
\newcommand*{\CHEP}{Center for High Energy Physics, Peking University, Beijing 100871, China}
\newcommand*{\ZZU}{School of Physics, Zhengzhou University, Zhengzhou 450001, China}

\author[a]{Hanlin Song}
\author[a,b,c]{Bo-Qiang Ma\corref{cor1}}

\address[a]{\PKU}
\address[b]{\CHEP}
\address[c]{\ZZU}
\cortext[cor1]{Corresponding author \ead{mabq@pku.edu.cn}}

\begin{abstract}
Lorentz invariance violation in photons can be quantified by measuring the difference in arrival times between high- and low-energy photons originating from gamma-ray bursts (GRBs). When analyzing data, it is crucial to consider the inherent time delay in the emission of these photons at the source of the GRB. In a recent study, three distinct models were evaluated to explain the intrinsic emission times of high-energy photons by analyzing 14 multi-GeV photon events detected from 8 GRBs using the Fermi Gamma-ray Space Telescope (FGST). In this study, we examine three remarkable GRB photons recorded by different observatories: the 99.3~GeV photon from GRB 221009A observed by FGST, the 1.07~TeV photon from GRB 190114C detected by the Major Atmospheric Gamma Imaging Cherenkov (MAGIC) telescope, and the 12.2~TeV photon from GRB 221009A observed by the Large High Altitude Air-shower Observatory (LHAASO). Our analysis indicates that the newly proposed model with a linear relationship between photon energy and intrinsic emission time can offer a consistent framework to explain the behavior of all three exceptional photons with a Lorentz violation scale $E_{\rm LV}\sim 3\times 10^{17}$~GeV.
\end{abstract}

\end{frontmatter}




Lorentz invariance, a fundamental principle of modern physics, posits that the laws of physics remain unchanged under Lorentz transformations. However, the possibility of Lorentz invariance violation (LV) has been a subject of intense theoretical and experimental scrutiny in recent years (for recent reviews, see, e.g. \cite{He:2022gyk,AlvesBatista:2023wqm}). Gamma-ray bursts (GRBs), as the most energetic and enigmatic events in the universe after the Big Bang, provide a unique laboratory for probing the fundamental properties of spacetime and particle interactions.

One intriguing avenue for investigating Lorentz invariance violation (LV) is through the analysis of high-energy photons emitted during gamma-ray bursts (GRBs). Discrepancies in the arrival times of high- and low-energy photons from GRBs may indicate potential deviations from Lorentz invariance, manifesting as time delays attributed to variations in photon velocities over vast cosmological distances~\cite{Amelino-Camelia:1996bln, Amelino-Camelia:1997ieq}. By examining these time delays and comparing them to theoretical frameworks, researchers can uncover insights into potential violations of Lorentz invariance and the intrinsic emission processes occurring at the sources of these astrophysical phenomena (for examples, see refs.~\cite{Xu:2016zxi,Xu:2016zsa}). Recent investigation such as the study by~\cite{plb138951}, has explored three distinct models to elucidate the intrinsic emission times of high-energy photons detected from multiple GRBs using data obtained from the Fermi Gamma-ray Space Telescope (FGST).

Building upon these findings, our study aims to apply the established methodology and outcomes from the previous researches~\cite{Xu:2016zxi,Xu:2016zsa,plb138951} to investigate Lorentz invariance violation in a new dataset of high-energy photon events from GRBs. We focus on testing Lorentz invariance violation by examining the arrival time differences of the respective highe-energy GRB photons from three observatories: the Fermi Gamma-ray Space Telescope (FGST)~\cite{Lesage:2023vvj}, the Major Atmospheric Gamma Imaging Cherenkov (MAGIC) telescope~ \cite{MAGIC:2019lau, MAGIC:2020egb}, and the Large High Altitude Air-shower Observatory (LHAASO)~\cite{LHAASO:2023lkv}. The results highlight the efficacy of a novel model (i.e., Model C in ref.~\cite{plb138951}) incorporating a linear relationship between photon energy and intrinsic emission time in explaining the behavior of these remarkable high-energy photons.

The study of gamma-ray bursts has a rich research history that has evolved over five decades.
Gamma-ray bursts were first discovered in the late 1960s by the Vela satellites, which were originally designed to detect nuclear tests in space. The initial observations of short, intense bursts of gamma rays sparked scientific interest and led to further investigations into the nature and origins of these mysterious events. Over the decades since their discovery, GRBs have been the subject of extensive research and study, with significant advancements in observational technology and theoretical models contributing to our understanding of these high-energy astrophysical phenomena.
Early satellite missions focused on exploring GRBs, laying the foundation for subsequent investigations. 
FGST played a pivotal role in advancing the study of GRBs by enabling the observation of GeV-range GRB photons. Subsequently, 
MAGIC contributed to the field by initiating research on TeV-range GRB photons. More recently, 
LHAASO has furthered the study by delving into multi-TeV GRB photons. These three observatories represent significant milestones in the exploration of different energy ranges of GRB photons, with FGST started on GeV photons, MAGIC on TeV photons, and LHAASO on multi-TeV photons. By examining three exemplary GRB photons detected by these observatories, we can trace the progression of research in the field of GRB studies and the increasing capabilities of both satellite and ground observations in capturing high-energy phenomena.
\new{The primary motivation for selecting the three ``remarkable GRB photons" is to establish a clear and coherent framework for understanding high-energy events within the context of our proposed approach. By focusing on these specific cases, we aim to demonstrate how our methodology can offer a self-consistent interpretation of new observations, a task that is often challenging with traditional analytical methods. To validate our findings and enhance our understanding of the implications of Lorentz invariance violation, further investigations will be essential. This includes exploring additional GRB events as well as other astrophysical phenomena, which will provide a more comprehensive context for our results.}

As an effective way to include the LV effect,
the dispersion relation for a photon with energy $E$ and momentum $p$ can be modified by the leading terms of Taylor series \cite{Jacob:2008bw,Xiao:2009xe} as:
\begin{equation}
E^2=p^2c^2\left[1-s_n\left(\frac{pc}{E_{\mathrm{LV},n}}\right)^n\right],
\end{equation}
where $s_n \equiv \pm 1$ is the indicator for high-energy photons traveling faster ($s_n = - 1$) or slower ($s_n = + 1$) than the low-energy photons and $E_{{\rm LV},n}$ denotes the $n$-th order energy scale of LV (with LV denoting Lorentz-invariance violation or light-speed variation) to be determined by the observation. 
By applying the relation $v = \partial E / \partial p$, the modified velocity is: 
\begin{equation}
v(E)=c\left[1-s_n\frac{n+1}{2}\left(\frac{pc}{E_{\mathrm{LV},n}}\right)^n\right].
\end{equation}

Taking into account the expansion of the universe, one can write the time delay caused by LV as \cite{Jacob:2008bw,Zhu:2022blp}:
\begin{equation}
\label{lorentzdelay}
\Delta t_{\mathrm{LV}}=s_{n}\frac{1+n}{2H_{0}}\frac{E_{\mathrm{h}}^{n}-E_{\rm l}^{n}}{E_{\mathrm{LV},n}^{n}}\int_{0}^{z}\frac{(1+z')^{n}\mathrm{d}z'}{\sqrt{\Omega_{\mathrm{m}}(1+z')^{3}+\Omega_{\Lambda}}},
\end{equation}
where $E_{\rm h}$ and $E_{\rm l}$ are energies of high- and low-energy photons, $z$ is the redshift of GRB, $H_0$ is the Hubble constant, and $\Omega_{\mathrm{m}}$ and $\Omega_{\Lambda}$ are matter density parameter and dark energy density parameter of the $\Lambda {\rm CDM}$ model. 
 
The observed time delay $\Delta t_{\mathrm{obs}}$ can be expressed by the LV time delay term in Eq.~\ref{lorentzdelay} and an intrinsic time delay term \cite{Ellis:2005sjy, Shao:2009bv, Zhang:2014wpb, Xu:2016zxi, Xu:2016zsa, Liu:2018qrg,Zhu:2021pml, Zhu:2021wtw, Zhu:2022usw, Huang:2019etr,plb138951}. Due to the expansion of the universe, it can be expressed as:
\begin{equation}
\label{obsdelay}
\Delta t_{\mathrm{obs}}=\Delta t_{\mathrm{LV}}+(1+z)\Delta t_{\mathrm{in}},
\end{equation}
where $\Delta t_{\mathrm{in}}$ is the intrinsic emission time delay between high- and low-energy photons in the GRB source frame.

In the earlier studies~\cite{Shao:2009bv, Zhang:2014wpb, Xu:2016zxi, Xu:2016zsa, Liu:2018qrg,Zhu:2021pml, Zhu:2021wtw, Zhu:2022usw}, the intrinsic emission time delay
$\Delta t_{\mathrm{in}}$ is treated as a common constant term in the analysis of Eq.~\ref{obsdelay} to extract the Lorentz violation scale.
As shown in \cite{Xu:2016zxi,Xu:2016zsa}, after analyzing  a dataset consisting of 14 multi-GeV photon events detected from 8 GRBs by the Fermi Gamma-ray Space Telescope (FGST),  the results suggest that $n=  1$, $s_n = + 1$, $E_{\rm LV,1} \simeq 3.60 \times 10^{17}$ GeV, and $\Delta t_{\rm in} = -10.7$ s for the mainline photons (see Fig.~2 of \cite{Xu:2016zsa}). These findings suggest that the subluminal aspect of cosmic photon Lorentz violation is permissible. 

It appears essential to consider the energy-dependence for the intrinsic emission time of photons in the GRB source frame~\cite{Wei:2016exb}. 
In a recently new study~\cite{plb138951}, a more general form for the intrinsic emission time with energy-dependence is introduced:
\begin{equation}
    \Delta t_{\rm{in}} = \Delta t_{\rm in, c} + \alpha E_{\rm s},
\end{equation}
where $\Delta t_{\rm in, c}$ is a common constant term and $E_{\rm s}$ is the source frame energy of high-energy photon with $\alpha$ being the coefficient. Then three distinct models in relating the observed time delay $\Delta t_{\mathrm{obs}}$ with the Lorentz violation time delay $\Delta t_{\mathrm{LV}}$ and the intrinsic time delay $\Delta t_{\rm in}$ are cataloged as:\\
\begin{itemize}
\item {\bf{Model A}}: with the inclusion of the Lorentz violation term $\Delta t_{\mathrm{LV}}$ by taking $\Delta t_{\rm in}$ as a common constant term for all high-energy photons under study
\begin{equation}
\label{obsdelayA}
\Delta t_{\mathrm{obs}}=\Delta t_{\mathrm{LV}}+(1+z)\Delta t_{\mathrm{in},c};
\end{equation}
\item {\bf{Model B}}: taking $\Delta t_{\rm in}$ as the general form with a common constant term and a linear-type energy dependence term without considering the Lorentz violation term
\begin{equation}
\label{obsdelayB}
\Delta t_{\mathrm{obs}}=(1+z)\Delta t_{\mathrm{in}};
\end{equation}
\item {\bf{Model C}}: taking $\Delta t_{\rm in}$ as the general form together with the Lorentz violation term 
\begin{equation}
\label{obsdelayC}
\Delta t_{\mathrm{obs}}=\Delta t_{\mathrm{LV}}+(1+z)\Delta t_{\mathrm{in}}.
\end{equation}
Actually, Model C combines Models A and B into a unified framework~\cite{plb138951}.
\end{itemize}

After applying the above three models to analyze the 14 multi-GeV photons from FGST by employing a comprehensive Bayesian parameter estimation approach for a multiple linear model~\cite{plb138951} with the \texttt{bilby} package~\cite{Ashton:2018jfp, Romero-Shaw:2020owr} for our computations, distinct scenarios about the Lorentz violation and intrinsic emission time are obtained and discussed in \cite{plb138951}.
It should be noted that excluding two short GRBs, 090510 and 140619B, from the dataset has negligible impact on the results. 
\new{It is reasonable to expect distinct behaviors in photon emissions between long GRBs and short GRBs, which necessitates a separate analysis of photon data from these two categories. However, given the limited number of high-energy photons available from short GRBs, we propose to defer a comprehensive analysis of these emissions to future studies, which may benefit from a larger sample of photon events from short GRBs.}

\begin{table*}[htbp]
  \centering
  \caption{The detailed information of remarkable high-energy GRB photons from FGST \cite{Lesage:2023vvj, Zhu:2022usw} for GRB 221009A, MAGIC for GRB 190114C \cite{MAGIC:2019lau, MAGIC:2020egb,Zhu:2021wtw}, and LHAASO for GRB 221009A \cite{LHAASO:2023lkv}. }
    \begin{tabular}{cccllll}
    \toprule
    Observatory & GRB   & $z$     & $E_{\rm h}$ (GeV)   &  $\frac{\Delta t_{\rm obs}}{1+z}$ (s)  \\
    \midrule
     FGST  & GRB 221009A & 0.151 & 99.3 $\pm \ 9.93$ &  -8.71 $\pm \ 4.35$ \\
     MAGIC & GRB 190114C & 0.4245 & 1070 $\pm \ 160.5 $ &  48.86 $\pm \ 3.51$ \\
    LHAASO & GRB 221009A & 0.151 & 12200 $\pm \ 2440$ &  340.19 $\pm \ 4.35$ \\
    \bottomrule
    \end{tabular}
  \label{GRB_data_table}
\end{table*}

\begin{figure*}[]
    \centering
    \begin{minipage}{0.45\textwidth}
        \centering
        \includegraphics[width=0.8\linewidth]{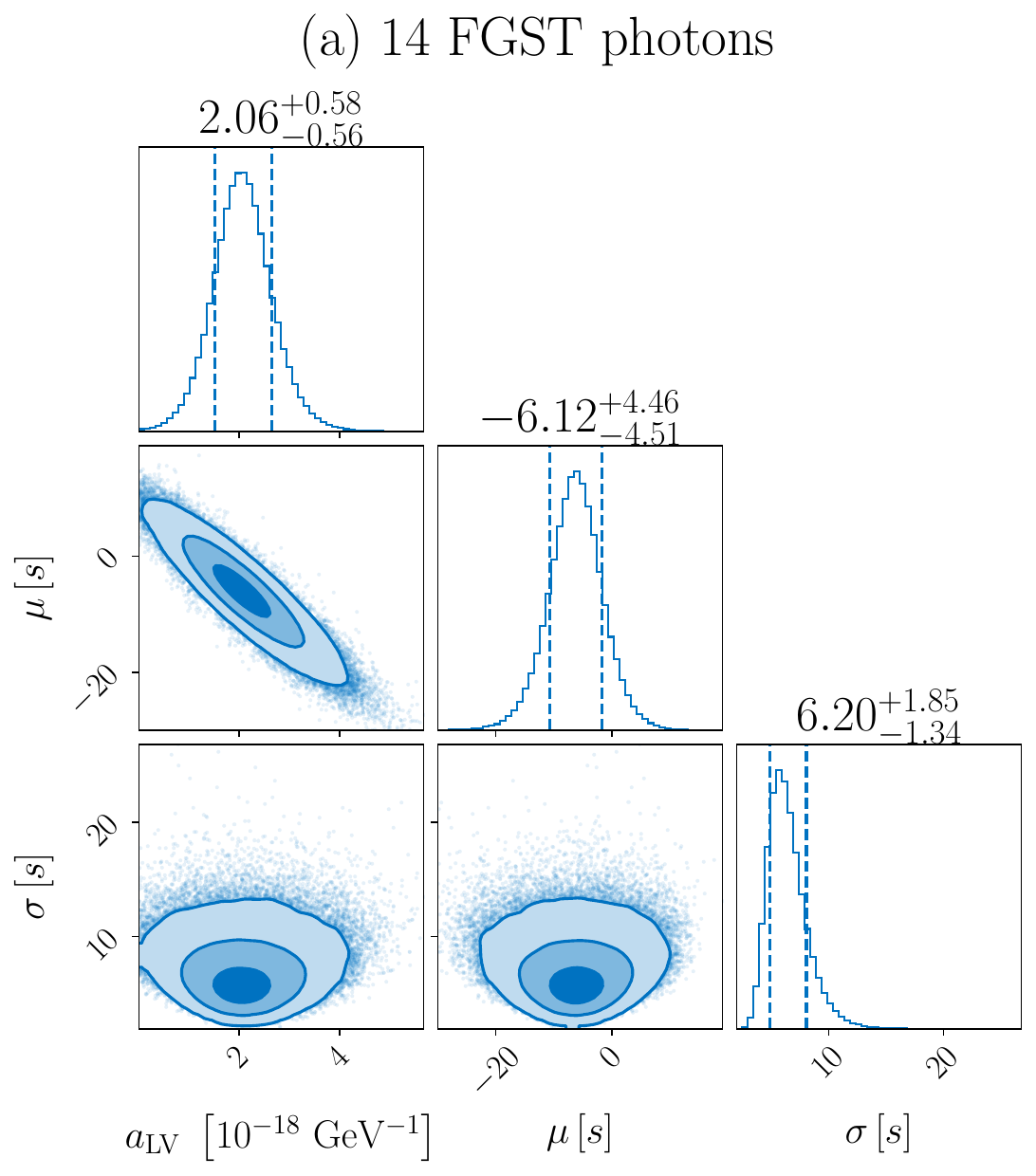}
    \end{minipage}\hfill
    \begin{minipage}{0.45\textwidth}
        \centering
        \includegraphics[width=0.8\linewidth]{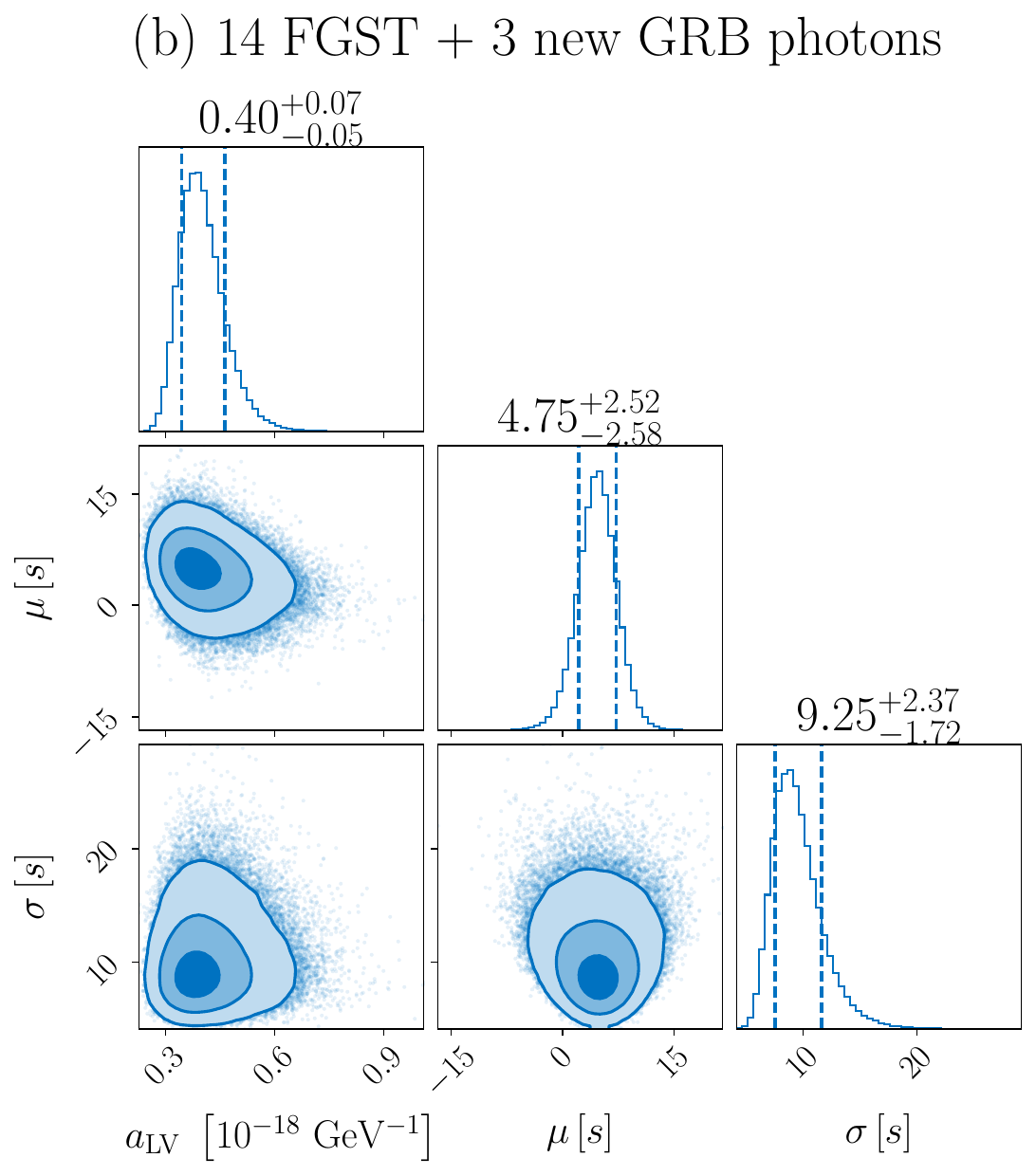}
    \end{minipage}
    
    \vspace{2em}

    \begin{minipage}{0.33\textwidth}
        \centering
        \includegraphics[width=0.95\linewidth]{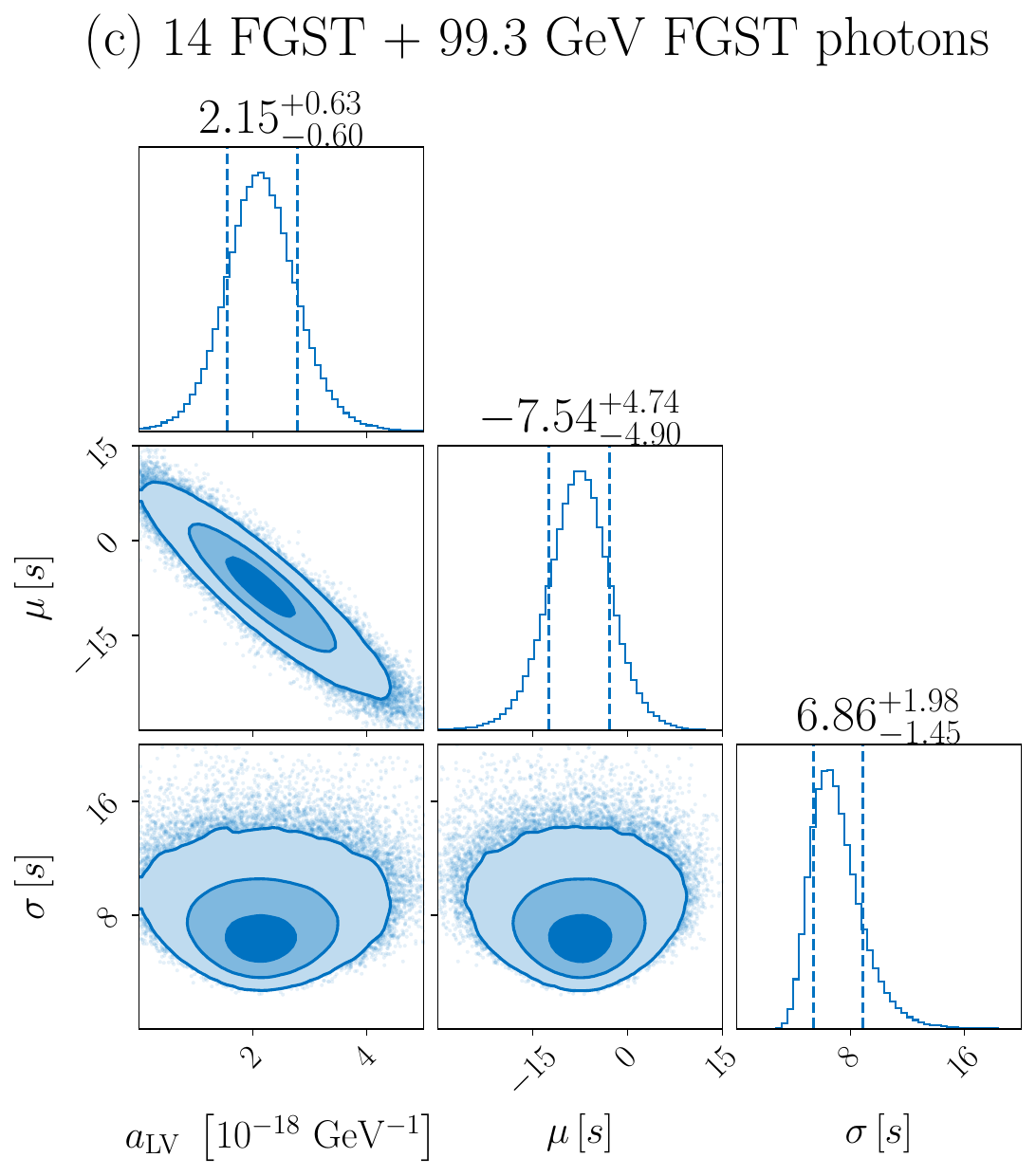}
    \end{minipage}\hfill
    \begin{minipage}{0.33\textwidth}
        \centering
        \includegraphics[width=0.95\linewidth]{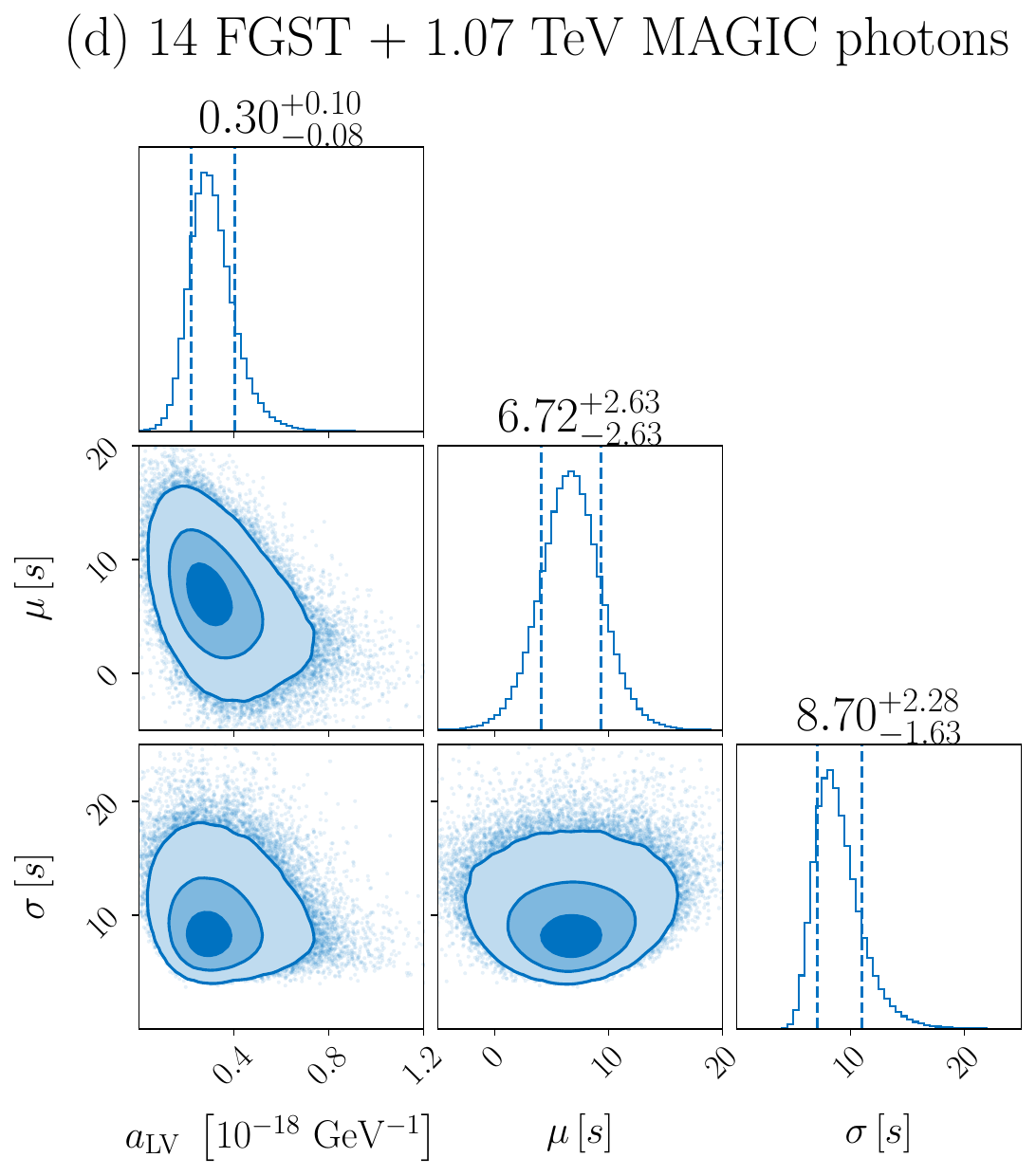}
    \end{minipage}\hfill
    \begin{minipage}{0.33\textwidth}
        \centering
        \includegraphics[width=0.95\linewidth]{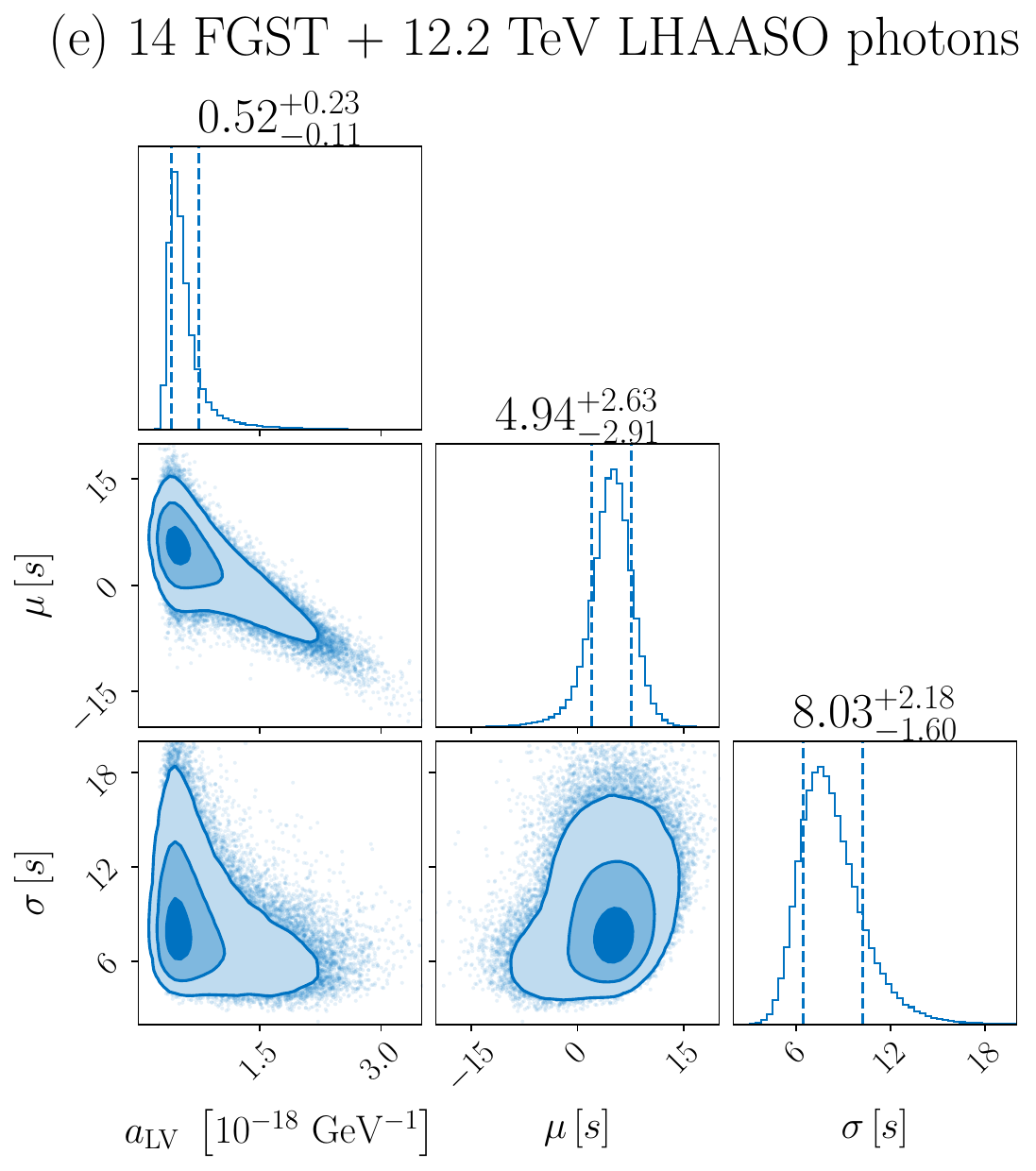}
    \end{minipage}

    \caption{The fitting results of five cases for Model A. Subfigure (a) shows the posterior distributions of fitting the 14 previous FGST photons in \cite{Xu:2016zsa, Xu:2016zxi}. Subfigure (b) shows the fitting results for the 14 previous FGST photons and three new remarkable photons observed by FGST, MAGIC and LHAASO. Subfigure (c) shows the fitting results for the 14 previous FGST photons and one new 99.3 GeV photon of GRB 221009A by FGST. Subfigure (d) shows the fitting results for  the 14 previous FGST photons and one new 1.07 TeV photon of GRB 190114C by MAGIC. Subfigure (e) shows the fitting results for the 14 previous FGST photons and one new 12.2 TeV photon of GRB 221009A by LHAASO. }
    \label{Model_A}
\end{figure*}

\begin{figure*}[]
    \centering
    \begin{minipage}{0.45\textwidth}
        \centering
        \includegraphics[width=0.8\linewidth]{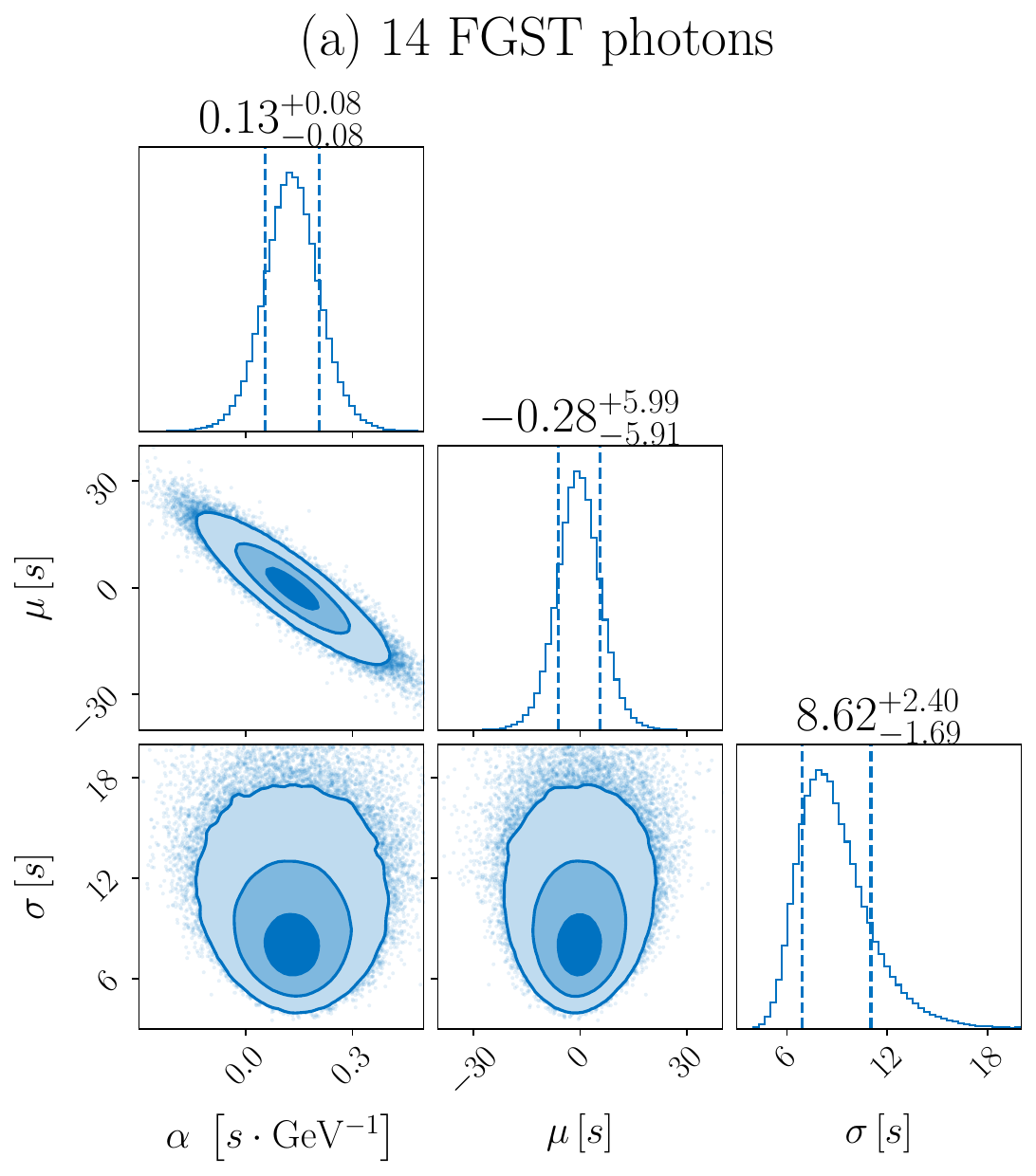}
    \end{minipage}\hfill
    \begin{minipage}{0.45\textwidth}
        \centering
        \includegraphics[width=0.8\linewidth]{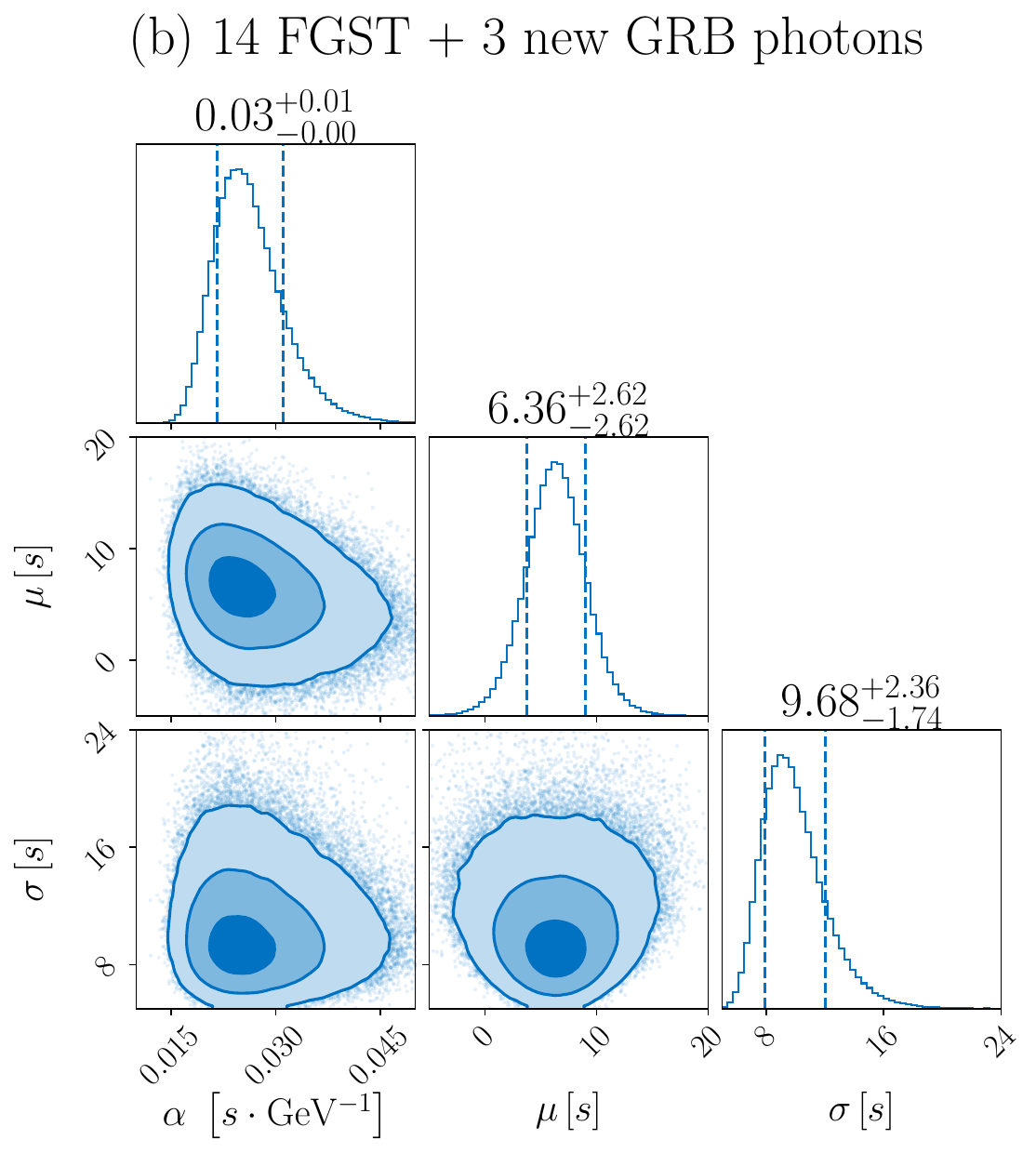}
    \end{minipage}
    
    \vspace{2em}

    \begin{minipage}{0.33\textwidth}
        \centering
        \includegraphics[width=0.95\linewidth]{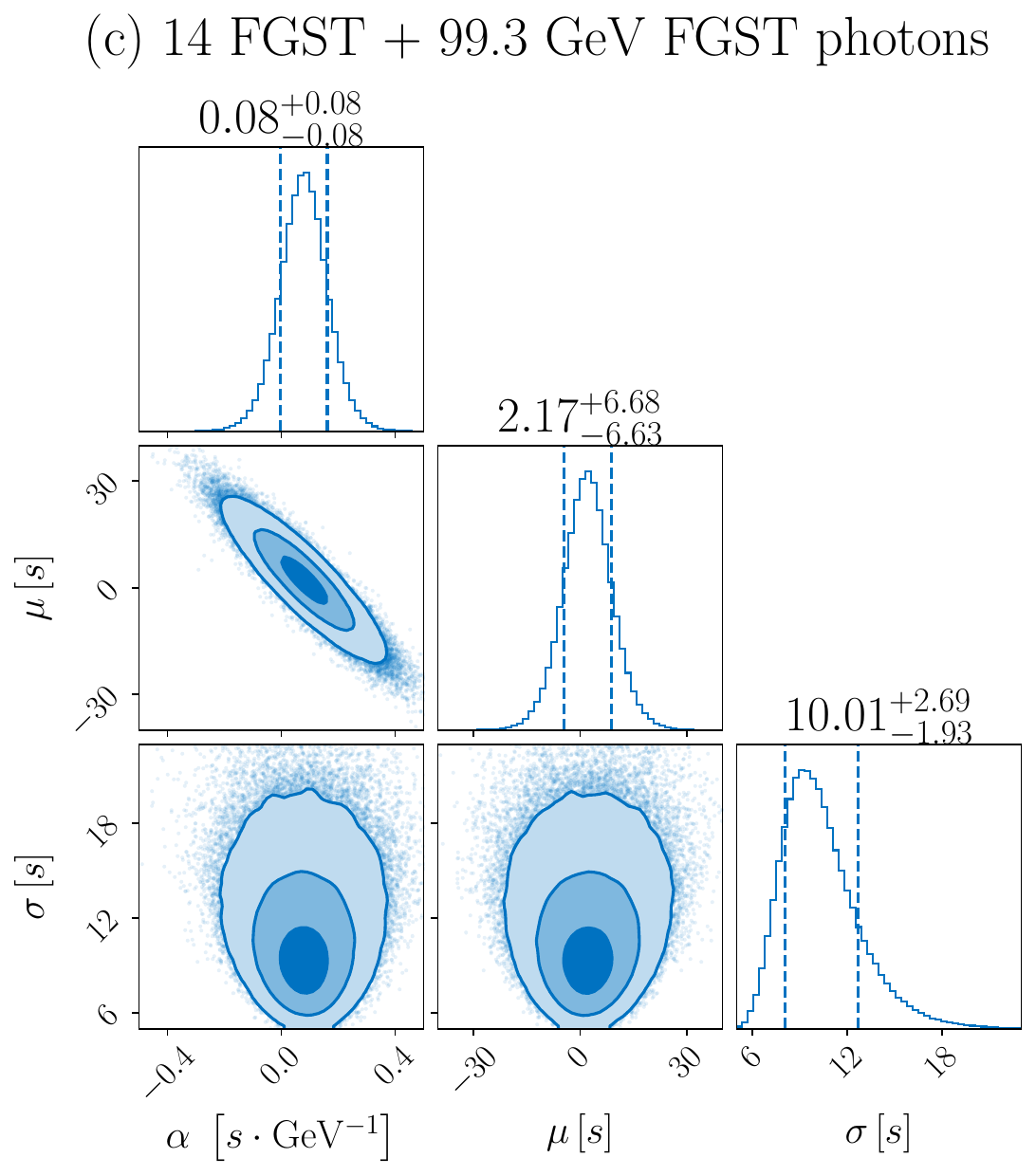}
    \end{minipage}\hfill
    \begin{minipage}{0.33\textwidth}
        \centering
        \includegraphics[width=0.95\linewidth]{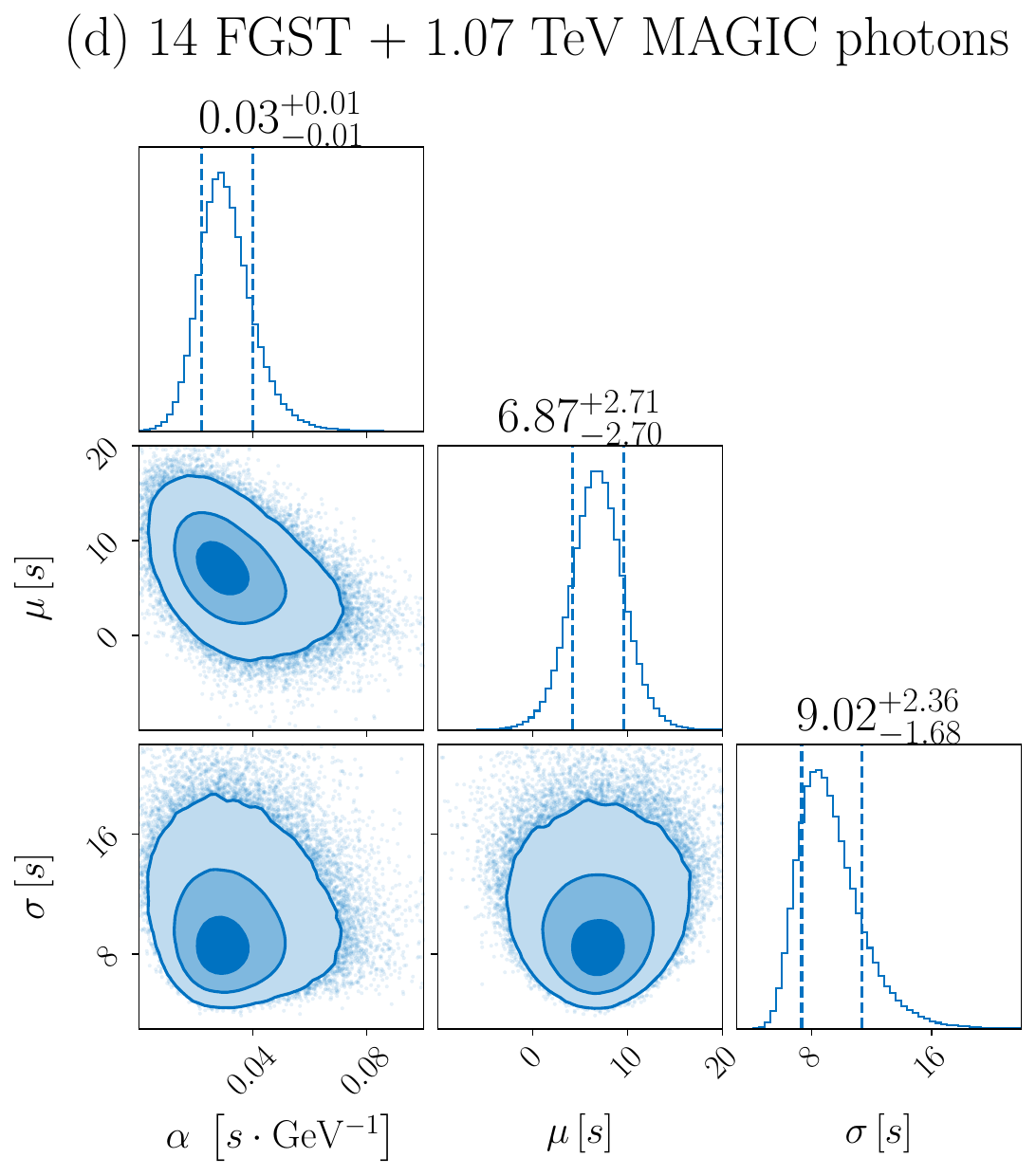}
    \end{minipage}\hfill
    \begin{minipage}{0.33\textwidth}
        \centering
        \includegraphics[width=0.95\linewidth]{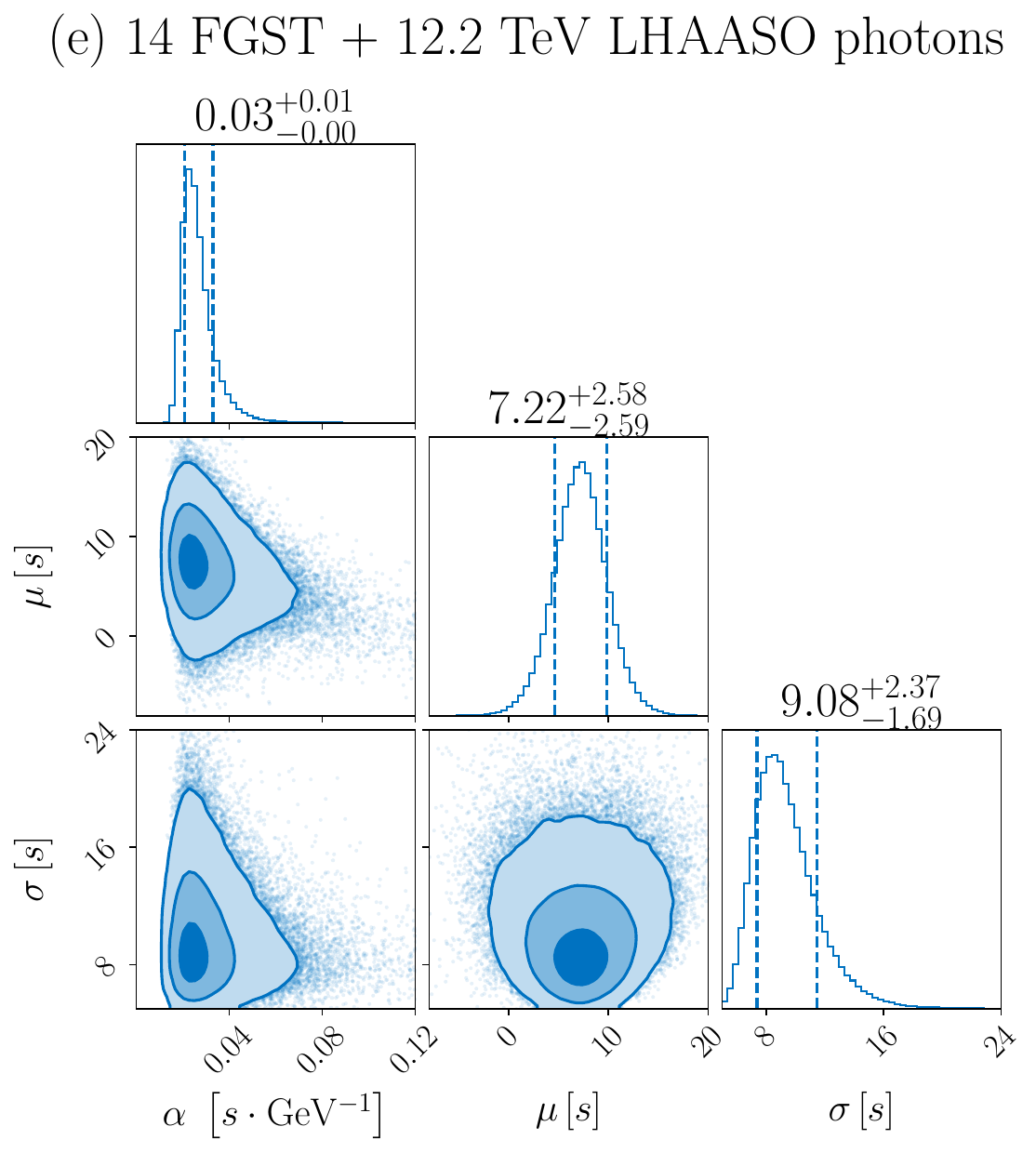}
    \end{minipage}

    \caption{Same as Fig.\ref{Model_A}, but for Model B.}
    \label{Model_B}
\end{figure*}

\begin{figure*}[]
    \centering
    \begin{minipage}{0.45\textwidth}
        \centering
        \includegraphics[width=0.8\linewidth]{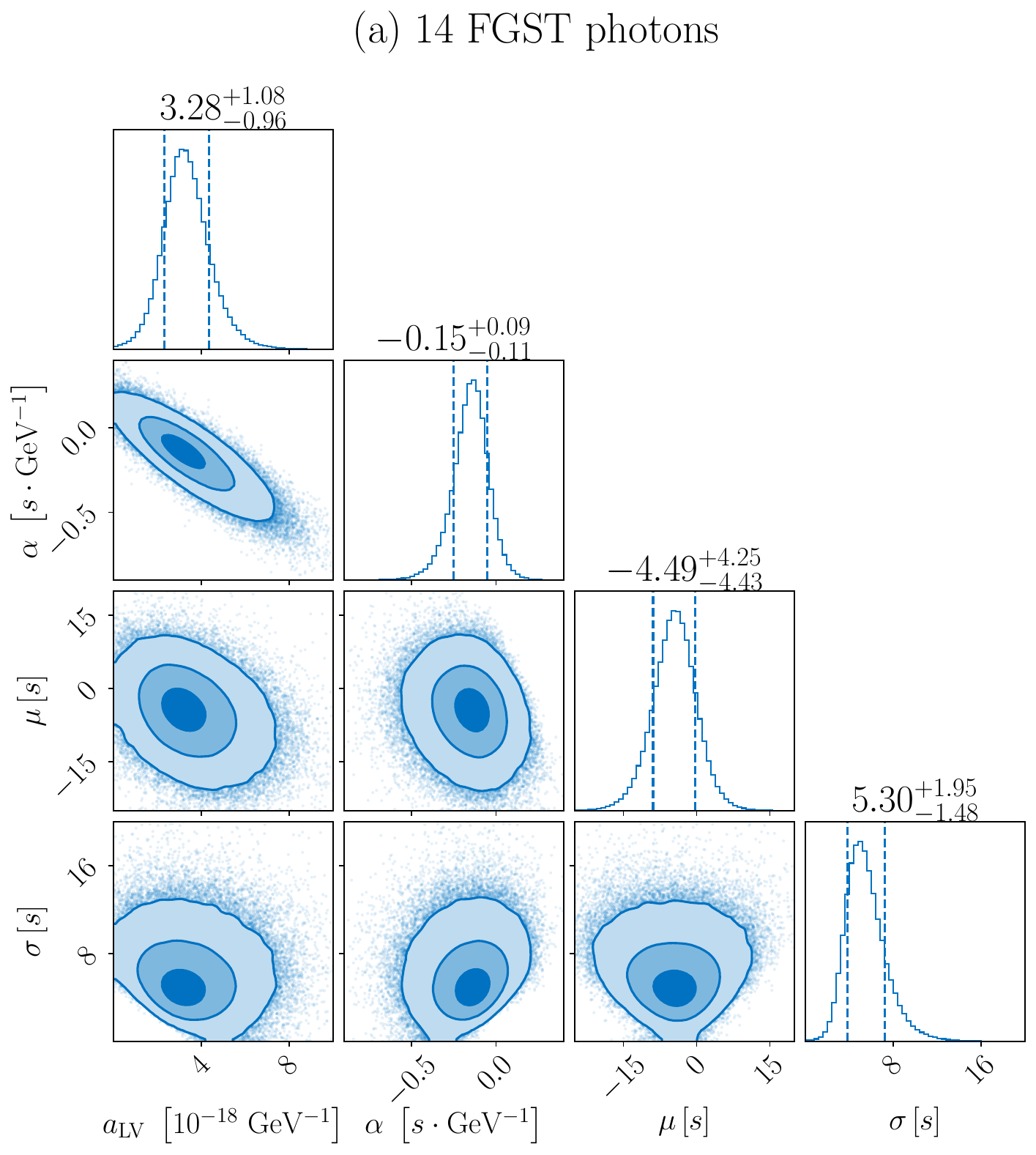}
    \end{minipage}\hfill
    \begin{minipage}{0.45\textwidth}
        \centering
        \includegraphics[width=0.8\linewidth]{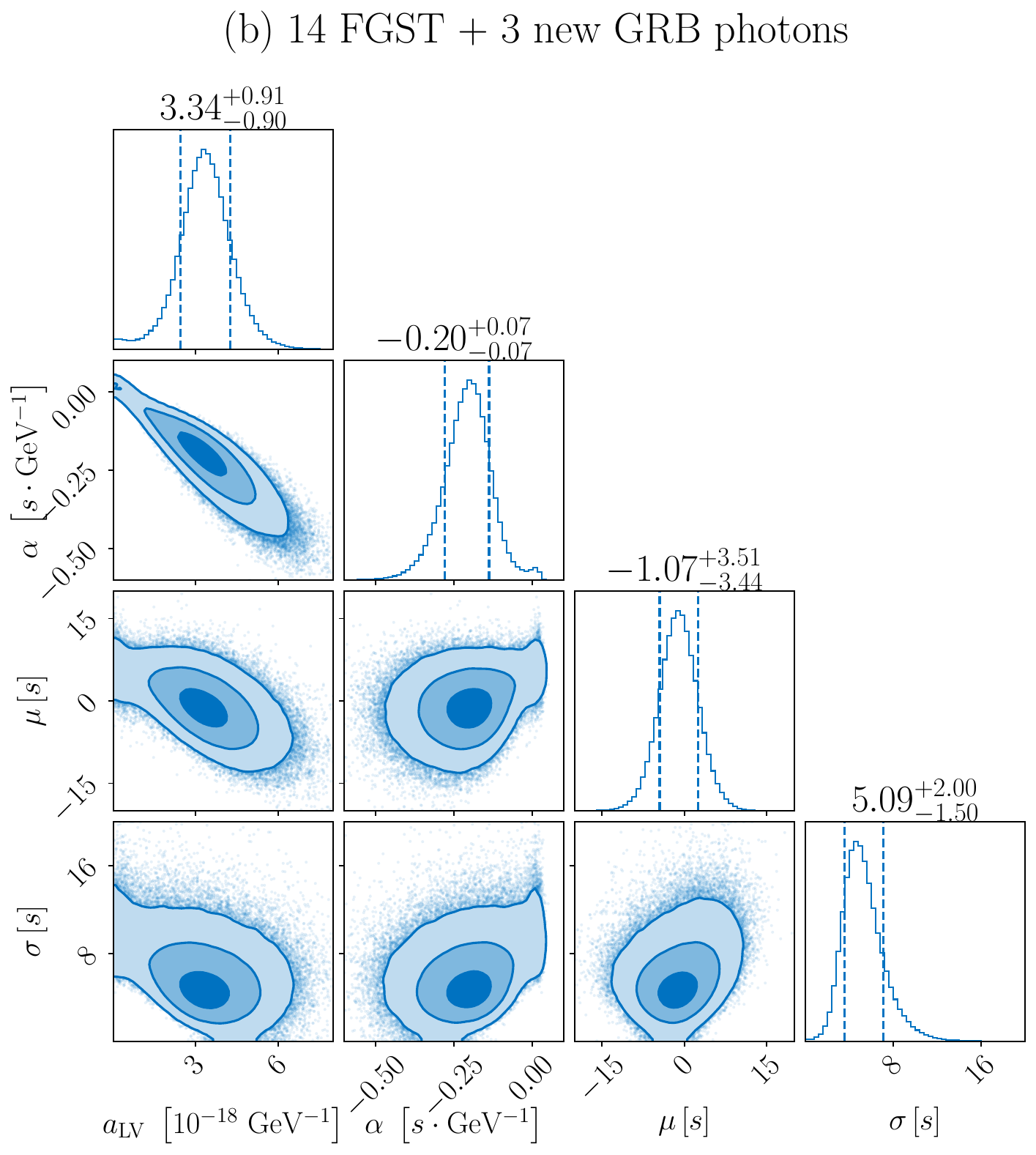}
    \end{minipage}
    
    \vspace{2em}

    \begin{minipage}{0.33\textwidth}
        \centering
        \includegraphics[width=0.95\linewidth]{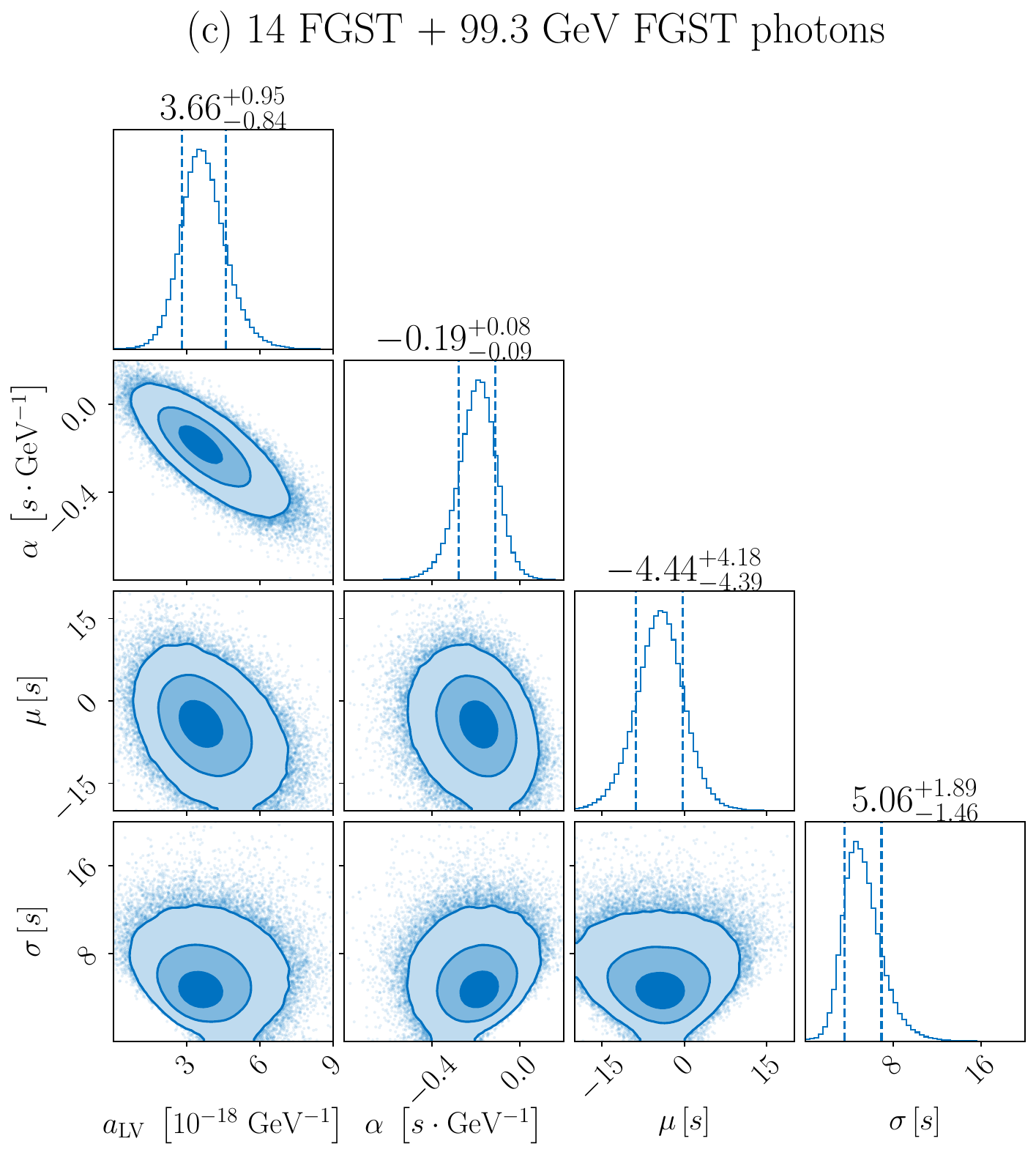}
    \end{minipage}\hfill
    \begin{minipage}{0.33\textwidth}
        \centering
        \includegraphics[width=0.95\linewidth]{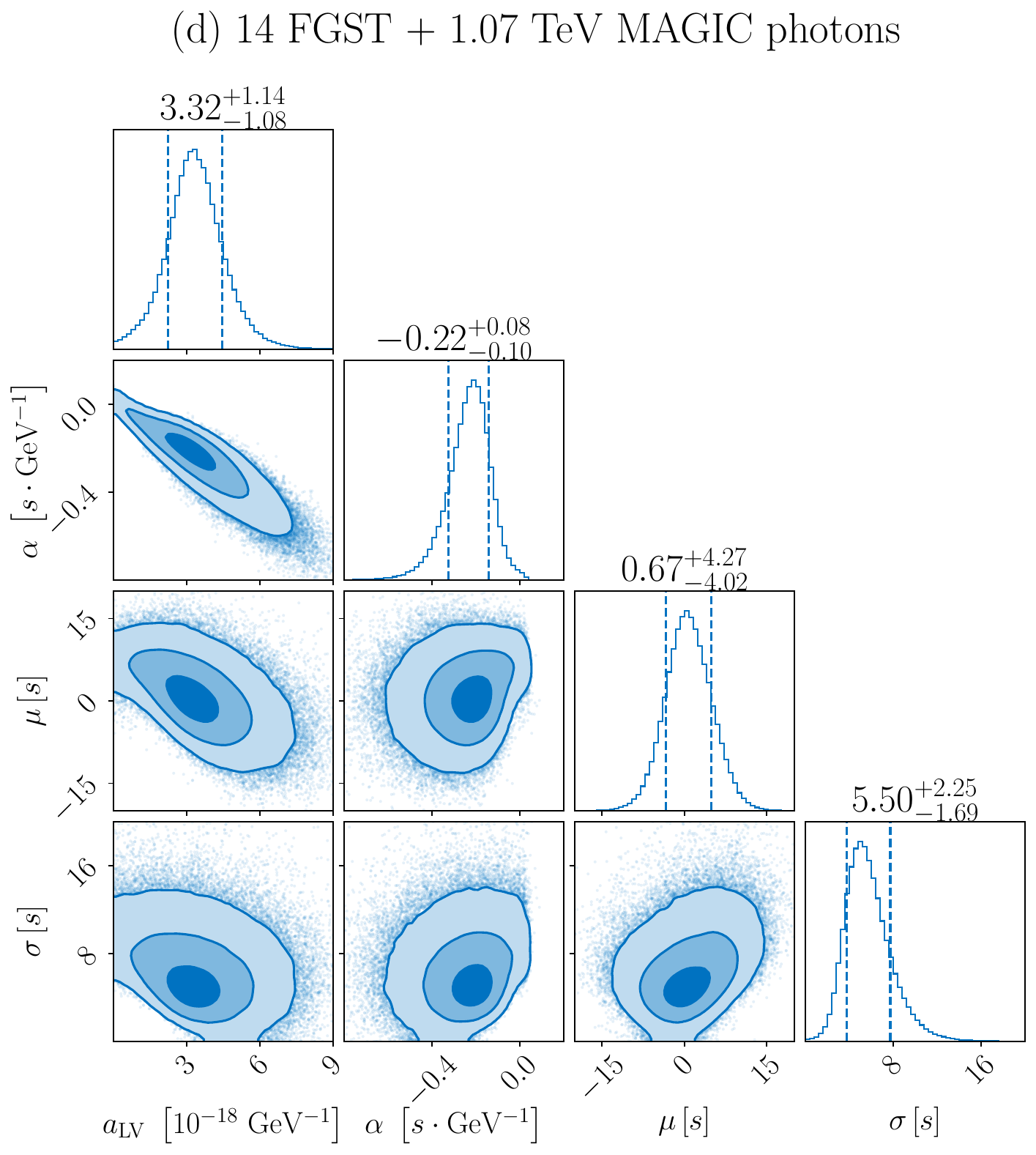}
    \end{minipage}\hfill
    \begin{minipage}{0.33\textwidth}
        \centering
        \includegraphics[width=0.95\linewidth]{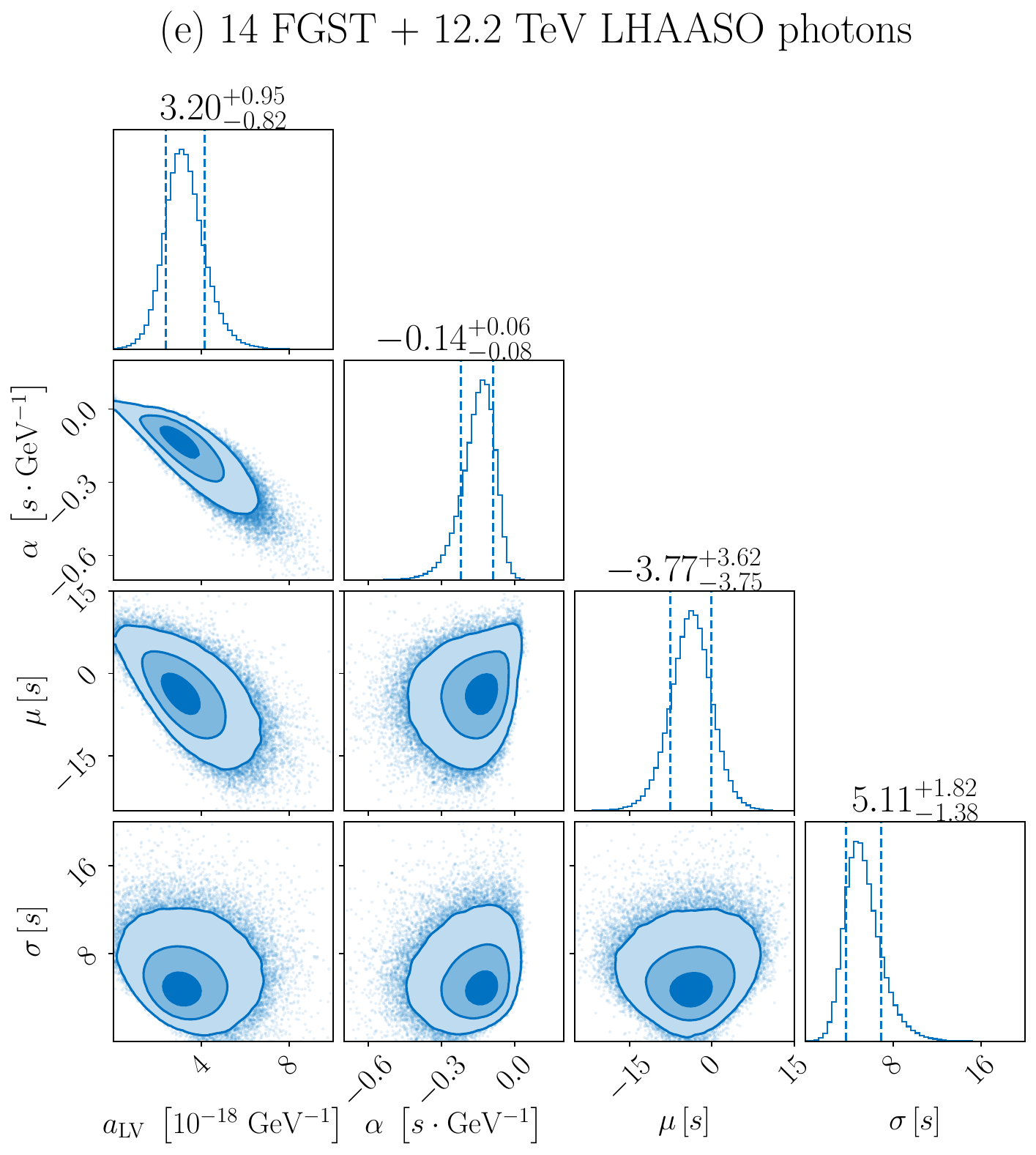}
    \end{minipage}

    \caption{Same as Fig.\ref{Model_A}, but for Model C.}
    \label{Model_C}
\end{figure*}

Now we shift our focus to include three specific high-energy photons originating from different GRBs: the 99.3 GeV photon from GRB 221009A observed by FGST \cite{Lesage:2023vvj}, the 1.07 TeV photon from GRB 190114C detected by MAGIC \cite{MAGIC:2019lau,MAGIC:2020egb}, and the 12.2 TeV photon from GRB 221009A observed by LHAASO \cite{LHAASO:2023lkv}. Detailed information about these three photons are listed in Table~\ref{GRB_data_table}. 
\new{In our data analysis, we need a reference time based on the first significant peak observed in the low-energy light curves from the Gamma-ray Burst Monitor (GBM) for each gamma-ray burst (GRB)~\cite{Xu:2016zxi, Xu:2016zsa, Liu:2018qrg}. The low-energy photon peaks for the initial eight GRBs align with those reported in previous studies \cite{Xu:2016zxi, Xu:2016zsa, Liu:2018qrg}. For the two new GRBs, GRB 190114C and GRB 221009A, we designate the low-energy peaks as $2.59 \pm 5.00$ seconds and  $251.33 \pm  5.00$ seconds, respectively, based on the Fermi-GBM light curves for each burst.}
The same as in \cite{plb138951}, we consider the common intrinsic time delay follows Gaussian distribution  $p\left(\Delta t_{\rm in,c} \right) \sim \mathcal{N} \left(\mu, \sigma\right)$. Meanwhile, the priors for each scenario are the same uniform distributions as in \cite{plb138951}. We also consider a $\pm 5$~s uncertainty for $\Delta t_{\rm obs}$ at observer frame for each high-energy photon events. We analyze 
the above GRB photons together with the dataset of 14 multi-GeV photons by FGST \cite{Xu:2016zsa, Xu:2016zxi} in three models with five different combinations of data sets for each model, and present the respective posterior distributions of the parameters $a_{\rm LV}=1/E_{\rm LV,1}=1/E_{\rm LV}$, $\alpha$, $\mu$ and $\sigma$ for each case in Fig.~\ref{Model_A} for Model A, Fig.~\ref{Model_B} for Model B, and Fig.~\ref{Model_C} for Model C.  

Five different cases of data combinations for the GRB photons under analysis are listed as followings:
\begin{itemize}
\item Case a: with only the 14 multi-GeV FGST photons same as \cite{Xu:2016zsa, Xu:2016zxi}, and with the fitting results of the corresponding posterior distributions presented in subfigure (a);
\item Case b: with the 14 multi-GeV FGST photons plus the 3 new remarkable GRB photons, and with the fitting results of the corresponding posterior distributions presented in subfigure (b);
\item Case c: with the 14 multi-GeV FGST photons plus the new 99.3~GeV FGST photon (which is the highest-energy GRB photon detected by FGST)~\cite{Lesage:2023vvj}, and with the fitting results of the corresponding posterior distributions presented in subfigure (c);
\item Case d: with the 14 multi-GeV FGST photons plus the new 1.07~TeV MAGIC photon (which is the only MAGIC released TeV event with known energy and arrival time)~\cite{MAGIC:2019lau, MAGIC:2020egb}, and with the fitting results of the corresponding posterior distributions presented in subfigure (d);
\item Case e: with the 14 multi-GeV FGST photons plus the new 12.2~TeV LHAASO photon (which is the highest-energy GRB photon ever detected)~\cite{LHAASO:2023lkv}, and with the fitting results of the corresponding posterior distributions presented in subfigure (e).
\end{itemize}

For the Model~A results presented in Fig.~\ref{Model_A}, the relevant physical parameters are the Lorentz violation scale $E_{\rm LV}=1/a_{\rm LV}$ and the intrinsic emission time constant $\mu$.  The estimated parameters are listed in Table~\ref{param_modelA}. We see that only Cases a and c (which only involve GeV-band photons) are consistent with each other with the Lorentz violation scale $E_{\rm LV}\sim 3\times 10^{17}$~GeV, whereas Cases b, d and e (which involve both GeV-band and TeV-band photons) with $E_{\rm LV}\sim 2\text{-}3\times 10^{18}$~GeV, differ significant from Cases a and c. Therefore Model A can not provide a consistent scenario to reconcile  the the remarkable MAGIC and LHAASO TeV photons with the multi-GeV photons for a same Lorentz violation scale $E_{\rm LV}$.

\begin{table*}[htbp]
  \centering
  \caption{Table of estimated parameters and the corresponding Lorentz violation scale $E_{\rm LV}$ for all five cases of Model A. }
    \begin{tabular}{ccccc}
    \toprule
    Case & $a_{\rm LV} ~ (10^{-18} ~ {\rm GeV}^{-1})$ & $\mu ~({\rm s})$ & $\sigma ~({\rm s})$ & $E_{\rm LV} ~(10^{17}~ {\rm GeV})$ \\
    \midrule
    Case a & $2.06^{+0.58}_{-0.56}$ & $-6.12^{+4.46}_{-4.51}$ & $6.20^{+1.85}_{-1.34}$ & $4.84^{+1.80}_{-1.06}$ \\
    Case b & $0.40^{+0.07}_{-0.05}$ & $4.75^{+2.52}_{-2.58}$  & $9.25^{+2.37}_{-1.72}$ & $25.27^{+3.85}_{-3.66}$ \\
    Case c & $2.15^{+0.63}_{-0.60}$ & $-7.54^{+4.74}_{-4.90}$ & $6.86^{+1.98}_{-1.45}$ & $4.65^{+1.80}_{-1.05}$  \\
    Case d & $0.30^{+0.10}_{-0.08}$ & $6.72^{+2.63}_{-2.63}$  & $8.70^{+2.28}_{-1.63}$ &$33.03^{+12.09}_{-8.27}$ \\
    Case e & $0.52^{+0.23}_{-0.11}$ & $4.94^{+2.63}_{-2.91}$  & $8.03^{+2.18}_{-1.60}$ & $19.26^{+5.30}_{-5.83}$ \\
    \bottomrule
    \end{tabular}%
  \label{param_modelA}%
\end{table*}

For the Model~B results presented in Fig.~\ref{Model_B}, the corresponding $E_{\rm LV}=\infty$ as there is no Lorentz violation term. The estimated parameters are listed in Table~\ref{param_modelB}. The corresponding physical parameters are the intrinsic emission time energy-dependent coefficient $\alpha$ and  the intrinsic emission time constant $\mu$. We see that  the central value  of $\alpha$ varies from 0.03 to 0.13 with weak consistence, whereas the central value  of $\mu$ is a small value from -0.44 to 10.19 second. There is no strong regulation for $\alpha$ but one can still draw a possible conclusion that the observed time delays of high-energy photons are mainly due to the intrinsic emission time delays at the GRB source frames by allowing a larger Lorentz violation scale $E_{\rm LV}$ comparable to or exceeding the Planck scale $E_{\rm p} \simeq 1.22 \times 10^{19}$~GeV.  In this situation we observe a scenario that  the high-energy photons are emitted subsequent to the low-energy photons at the GRB source frames.

\begin{table*}[]
  \centering
  \caption{Table of estimated parameters for all five cases of Model B. }
    \begin{tabular}{ccccc}
    \toprule
    Case & $\alpha ~( \rm{s} \cdot {\rm GeV}^{-1})$ & $\mu ~({\rm s})$ & $\sigma ~({\rm s})$ \\
    \midrule
    Case a & $0.13^{+0.08}_{-0.08}$ & $-0.28^{+5.99}_{-5.91}$ & $8.62^{+2.40}_{-1.69}$  \\
    Case b & $0.03^{+0.01}_{-0.00}$ & $6.36^{+2.62}_{-2.62}$  & $9.68^{+2.36}_{-1.74}$  \\
    Case c & $0.08^{+0.08}_{-0.08}$ & $2.17^{+6.68}_{-6.63}$ & $10.01^{+2.69}_{-1.93}$   \\
    Case d & $0.03^{+0.01}_{-0.01}$ & $6.87^{+2.71}_{-2.70}$  & $9.02^{+2.36}_{-1.68}$  \\
    Case e & $0.03^{+0.01}_{-0.00}$ & $7.22^{+2.58}_{-2.59}$  & $9.08^{+2.37}_{-1.69}$  \\
    \bottomrule
    \end{tabular}%
  \label{param_modelB}%
\end{table*}

For the Model~C results presented in Fig.~\ref{Model_C},
the relevant physical parameters are the Lorentz violation scale $E_{\rm LV}$, the intrinsic emission time energy-dependent coefficient $\alpha$, and the intrinsic emission time constant $\mu$. The estimated parameters are listed in Table~\ref{param_modelC}. It is a surprise that the three parameters for all of the five cases are consistent with each other for $E_{\rm LV}\sim 3\times 10^{17}$~GeV,
$\alpha \sim -0.2$~s$\cdot$GeV$^{-1}$, and $\mu \sim 0$ second. We obtain a physics scenario similar to that of previous studies~\cite{Xu:2016zsa,Xu:2016zxi,plb138951}, that the results support the Lorentz violation with the violation scale $E_{\rm LV}\sim 3\times 10^{17}$~GeV and the high-energy photons are emitted prior to the low-energy photons at the GRB source frames. More detailed illustrations and in-depth analyses on the results from Model~C can be found in Ref.~\cite{song_and_ma_ApJ}.

\begin{table*}[]
  \centering
  \caption{Table of estimated parameters and the corresponding Lorentz violation scale $E_{\rm LV}$ for all five cases of Model C. }
    \begin{tabular}{cccccc}
    \toprule
    Case & $a_{\rm LV} ~ (10^{-18} ~ {\rm GeV}^{-1})$ & $\alpha ~( \rm{s} \cdot {\rm GeV}^{-1}1) $ & $\mu ~({\rm s})$ & $\sigma ~({\rm s})$ & $E_{\rm LV} ~(10^{17}~ {\rm GeV})$ \\
    \midrule
    Case a & $3.28^{+1.08}_{-0.96}$ & $-0.15^{+0.09}_{-0.11}$ & $-4.49^{+4.25}_{-4.43}$ & $5.30^{+1.95}_{-1.48}$ & $3.05^{+1.25}_{-0.76}$ \\
    Case b & $3.34^{+0.91}_{-0.90}$ & $-0.20^{+0.07}_{-0.07}$  & $-1.07^{+3.51}_{-3.44}$ & $5.09^{+2.00}_{-1.50}$ & $3.00^{+1.11}_{-0.64}$ \\
    Case c & $3.66^{+0.95}_{-0.84}$ & $-0.19^{+0.08}_{-0.09}$ & $-4.44^{+4.18}_{-4.39}$ & $5.06^{+1.89}_{-1.46}$ & $2.74^{+0.81}_{-0.57}$  \\
    Case d & $3.32^{+1.14}_{-1.08}$ & $-0.22^{+0.08}_{-0.10}$  & $0.67^{+4.27}_{-4.02}$ & $5.50^{+2.25}_{-1.69}$ &$3.01^{+1.45}_{-0.77}$ \\
    Case e & $3.20^{+0.95}_{-0.82}$ & $-0.14^{+0.06}_{-0.08}$  & $-3.77^{+3.62}_{-3.75}$ & $5.11^{+1.82}_{-1.38}$ & $3.13^{+1.08}_{-0.71}$ \\
    \bottomrule
    \end{tabular}%
  \label{param_modelC}%
\end{table*}

From a technical perspective, increasing the number of free parameters does not always enhance the performance of a physics model. Fortunately, one can utilize the Akaike information criterion for model selection to determine the model that best fits the data \cite{akaike1981likelihood, Biesiada:2009zz}. The Akaike criterion is based on the Kullback-Leibler information, which quantifies the information loss between the true physical mechanism and the effective model. For a model with likelihood $\mathcal{L}$ and $k$ 
free parameters, the Akaike information criterion (AIC) is given by \cite{akaike1981likelihood, Biesiada:2009zz}, 
\begin{equation} 
{\rm AIC} = 2k - 2\ln{\mathcal{L}(\hat{\theta} \mid {\rm data})}, 
\end{equation} 
where $\hat{\theta}$ represents the best-fit parameter obtained through maximum likelihood estimation. A lower value of ${\rm AIC}$ indicates better performance of the model.

A more useful quantity can be defined as \cite{akaike1981likelihood, Biesiada:2009zz},
\begin{equation} 
\Delta_{i} = {\rm AIC_i} - {\rm AIC_{\rm min}}, 
\end{equation} 
where ${\rm AIC_i}$ denotes the ${\rm AIC}$ value for each model and ${\rm AIC_{\rm min}}$ denotes the minimum value of ${\rm AIC_i}$ among all models need to be compared. \new{The ${\rm AIC}$ values for each model and each case are shown in Table~\ref{aic_table_individual}.} The $\Delta_{i}$ for all five cases with three models used in this work are listed in Table~\ref{aic_table}. 
It is evident that Model C exhibits the best performance, whereas Model B demonstrates the poorest performance.

\begin{table}[htbp]
  \centering
  \caption{Values of ${\rm AIC}$ for all five cases with three models.}
    \begin{tabular}{lrrrrr}
    \toprule
    \textbf{Model} & \textbf{Case a} &\textbf{Case b} & \textbf{Case c} & \textbf{Case d} &\textbf{Case e} \\
    \midrule
    Model A & 69.63  & 101.46  & 77.91  & 83.82  & 86.33  \\
    Model B & 77.76  & 102.97  & 87.75  & 84.69  & 89.69  \\
    Model C & 69.09  & 100.87  & 73.89  & 82.55  & 81.90  \\
    \bottomrule
    \end{tabular}%
  \label{aic_table_individual}%
\end{table}%

\begin{table}[htbp]
  \centering
  \caption{Values of $\Delta_{i}$ for all five cases with three models.}
    \begin{tabular}{lrrrrr}
    \toprule
    \textbf{Model} & \textbf{Case a} &\textbf{Case b} & \textbf{Case c} & \textbf{Case d} &\textbf{Case e} \\
    \midrule
    Model A & 0.54  & 0.59  & 4.02  & 1.27  & 4.43  \\
    Model B & 8.67  & 2.10  & 13.86  & 2.15  & 7.80  \\
    Model C & 0.00  & 0.00  & 0.00  & 0.00  & 0.00  \\
    \bottomrule
    \end{tabular}%
  \label{aic_table}%
\end{table}%

Our analysis of the arrival time differences between high- and low-energy photons from these GRBs reveals intriguing patterns that suggest the presence of Lorentz invariance violation. By comparing the observed data to different models for intrinsic emission times of high-energy photons, we find that the newly proposed model with a linear dependence on photon energy (i.e., Model C~\cite{plb138951}) provides a consistent explanation for the behavior of these remarkable photons. These results offer valuable insights into the potential violation of Lorentz invariance in the context of extreme astrophysical phenomena.

The constraints on the Lorentz violation scale $E_{\rm{LV}}$
present notable variability across the existing literature. The discrepancies in these results can be attributed to the diverse methodologies employed in analyzing different datasets under varying assumptions. This variability is exemplified by the significant difference in the scenarios concerning the intrinsic emission times of high-energy photons at the source of gamma-ray bursts (GRBs) between Model B and Model C. Model C proposes a scenario in which high-energy photons are emitted earlier at the GRB source, with a Lorentz violation scale 
$E_{\rm LV}\sim 3\times 10^{17}$~GeV. In contrast, Model B suggests a scenario where high-energy photons are emitted later at the GRB source, with a substantial Lorentz violation scale $E_{\rm LV}$ comparable to or exceeding 
the Planck scale $E_{\rm p} \sim 10^{19}$~GeV.

Upon closer examination, it becomes evident that the scenario in Model B is based on specific assumptions, whereas the scenario in Model C is derived purely through data fitting without arbitrary preconceptions.

\new{The discussions above may elucidate why our scenario in Model C can yield signals indicative of Lorentz violation, whereas other studies report negative results~\cite{FermiGBMLAT:2009nfe,Xiao:2009xe,MAGIC:2020egb,LHAASO:2024lub}. One key reason is that those analyses typically rely on photons emitted solely from a single gamma-ray burst, often accompanied by additional assumptions, such as the premise that high-energy photons are emitted no earlier than low-energy photons from the GRB source. In contrast, our analysis incorporates the energy dependence of emission times for photons in the GRB source frame, utilizing model parameters derived exclusively from fitting GRB photons across multiple GRBs with varying redshifts. This comprehensive approach allows us to capture a broader range of emission dynamics, thereby enhancing our sensitivity to potential signals of Lorentz violation.}

Regarding possible concerns regarding the astrophysical interpretations to elucidate the data, it is important to note that conventional GRB models have traditionally been developed based on observations of GRB photons without considering Lorentz violation. If our suggested scenario indeed indicates the presence of Lorentz violation, it becomes imperative to reconstruct the GRB model incorporating this new scenario rather than dismissing Lorentz violation by relying on pre-existing GRB models that were formulated without considering such violations.

For instance, our prior investigations~\cite{Xu:2016zxi,Xu:2016zsa,plb138951}, including the current study, have uncovered a scenario where high-energy photons are emitted before low-energy photons at the GRB source, a phenomenon deduced solely through data fitting. This anticipation of a preburst phase for high-energy photons has been previously explored in studies such as~\cite{Zhu:2021pml,Chen:2019avc,Zhu:2021wtw}, yielding promising indications. 
\new{From a phenomenological perspective, a recent survey~\cite{Liu:2024qbt} of the GRB 221009A data has revealed direct signals indicating a preburst stage of GRBs, characterized by the earlier emission of higher-energy photons. Theoretically, we can anticipate a cooling and expanding process in GRBs that is analogous to the dynamics of the Big Bang. This leads to a scenario in which high-energy emissions are generated prior to the complete realization of lower-energy emissions. }

\new{In addition to the time delay analysis of Lorentz invariance violation (LV) from GRB photons, there are potential signals arising from threshold anomalies caused by LV in the interactions of high-energy GRB photons with the extragalactic background light (EBL). Recent literature has explored features related to LV, particularly in connection with the remarkable multi-TeV photons detected by LHAASO~\citep{LHAASO:2023lkv}. Notably, it has been suggested that the condition \(E_{\mathrm{LV}} < 0.1 E_{\mathrm{P}}\) provides a viable explanation for the LHAASO results concerning EBL interactions \cite{Li:2022wxc, Li:2023rgc, Li:2023rhj, LHAASO:2023lkv}. This perspective aligns with the conclusions drawn in the present study based on our time delay analysis.
From another viewpoint, the features associated with LV in the interaction of high-energy photons with the EBL are model-dependent and yield varying predictions. For example, within the framework of doubly special relativity~\cite{Amelino-Camelia:2002cqb, Amelino-Camelia:2002uql, Amelino-Camelia:2000stu}, only a threshold shift is anticipated, with no additional anomalies. Consequently, the LV features related to EBL interactions can primarily serve to constrain specific LV models rather than provide a comprehensive assessment of the overall effects of LV.}

The implications of our findings extend beyond the realm of astrophysics, potentially challenging our understanding of fundamental symmetries in nature. The detection of Lorentz invariance violation in the emission of high-energy photons from GRBs opens up new avenues for exploring the underlying physics of these cosmic events. Further research is needed to corroborate these results and delve deeper into the implications of Lorentz invariance violation in the context of high-energy astrophysics.

\new{One significant issue pertains to the potential implications of Lorentz invariance violation (LV) on atmospheric shower development and gamma-ray energy measurements. If LV is indeed present, it could substantially influence the dynamics of high-energy particle interactions within the atmosphere, leading to observable effects on the characteristics of the resulting particle showers.
LV may affect both the energy and timing of gamma-ray emissions, complicating the interpretation of energy distributions in detected gamma-ray signals. The atmospheric medium, unlike a vacuum, introduces additional complexities such as scattering and absorption, which could either mask or amplify the effects of LV.
For example, high-energy particles traveling through the atmosphere at speeds exceeding the speed of light in that medium generate Cherenkov radiation, which can be detected by ground-based observatories. The presence of LV may alter the conditions under which this radiation is produced, potentially resulting in deviations from the expected Cherenkov radiation patterns.
This scenario presents a unique opportunity to probe the effects of LV indirectly through the analysis of Cherenkov radiation. By investigating these deviations, we can gain valuable insights into the nature of LV and its implications for high-energy astrophysical processes.
Moreover, the predicted features arising from LV may vary across different theoretical frameworks, suggesting that the effects of LV are not uniform. Thus, exploring the interplay between LV and atmospheric interactions not only deepens our understanding of fundamental physics but also opens new avenues for research in high-energy astrophysics.}

\new{Our analysis is based on a limited number of events, specifically three new high-energy photons in addition to previously studied cases. The rationale for concentrating on these particular events is to establish a coherent framework for understanding high-energy emissions and to systematically demonstrate the potential for identifying time delays of photons from GRBs across different redshifts. However, we acknowledge that expanding our dataset would significantly enhance the robustness of our findings and address concerns related to selection biases. We are actively progressing in the analysis of additional datasets from Fermi-LAT and LHAASO, and we anticipate that the outcomes will align with the conclusions presented in this manuscript, as they are derived using the same analytical methods.}

Through the application of the established models and methodologies to new data, we endeavor to deepen our understanding of Lorentz invariance violation and its implications in the context of high-energy astrophysical phenomena. By extending the analysis to encompass a wider array of GRB observations, we strive to validate the proposed model and its ability to provide a comprehensive framework for interpreting the emission characteristics of high-energy photons from diverse astrophysical sources. This research contributes to the ongoing exploration of fundamental physics principles in the realm of gamma-ray bursts and provides valuable insights into the behavior of high-energy photons in extreme astrophysical environments.
\\

{\bf{Acknowledgements:}}
This work is supported by National Natural Science Foundation of China under Grants No.~12335006 and No.~12075003. This work is also supported by High-performance Computing Platform of Peking University.

\bibliographystyle{elsarticle-num}
\bibliography{scibib}

\end{document}